\documentstyle[preprint,aps]{revtex}
\begin{document}
\draft
\def\bra#1{{\langle #1{\left| \right.}}}
\def\ket#1{{{\left.\right|} #1\rangle}}
\def\bfgreek#1{ \mbox{\boldmath$#1$}}
\title{A chiral bag model approach to delta electroproduction}
\author{D. H. Lu, A. W. Thomas, and  A. G. Williams}
\address{Department of Physics and Mathematical Physics,
	University of Adelaide, Australia 5005\\
	and\\
	 Institute for Theoretical Physics, 
	University of Adelaide, Australia 5005}
\maketitle
\begin{abstract}

Helicity amplitudes for the $\gamma^* N \rightarrow\Delta$ transition 
are calculated using the cloudy bag model.
A correction for center-of-mass motion is carried out
using a modified  Peierls-Thouless projection method. This reduces the 
magnitude of the transition amplitudes at small momentum transfer and 
enhances them at modest momentum transfers.
Our calculation shows that the pion cloud contributes substantially
to the transition helicity amplitudes, with the final result giving
reasonable agreement with the corresponding experimental values.
\end{abstract}

\section{Introduction}

The nucleon-delta electromagnetic transition amplitude is
an outstanding example of the success of the quark model\cite{CLOSE}.
There have been many theoretical and experimental explorations 
of this transition process. 
In a naive quark model the $\gamma^* N\rightarrow\Delta$ 
transition occurs only 
by an M1 transition, while the E2 process is fully suppressed\cite{BECCHI65}.
In  more sophisticated models, quarks can interact through, for example,
 one-gluon 
exchange in addition to the confinement potential between them. Then it is 
possible for configuration mixing, involving the excitation of one quark to 
a $d$-state, to
generate a  small, but nonvanishing, E2 amplitude\cite{IK82}.
To extract the $\gamma^* N\rightarrow\Delta$ amplitude from experimental data 
is  not an easy task. 
There are some uncertainties on the subtraction of  background,
 and the results are somewhat model dependent\cite{DAVID91}. 
With the advent of new generation of accelerators,
much more accurate measurements will be made. 
The prospect of the high quality
data  should test various hadron models well 
and help to build  more realistic ones.

The cloudy  bag model (CBM)\cite{CBM} improves the MIT bag model\cite{MIT}
 by introducing an elementary, perturbative pion field which 
couples to quarks in the bag in such a way 
that chiral symmetry is restored.
The  pion field significantly improves the predictions of the static 
 properties of baryons. It  also provides a   convenient connection to 
the traditional language in nuclear physics.
Kalbermann and Eisenberg (KE)\cite{KE83} first  used the cloudy bag 
model to calculate the photoproduction of delta and Roper resonances 
at the threshold. They showed that the pionic corrections can 
substantially improve the predictions of the helicity amplitudes. 
Subsequently Bermuth et al.\cite{Bermuth86} updated the KE calculation with
some algebraic corrections and  the appropriate normalization
of baryon states.
By comparing pseudoscalar (PS) and pseudovector (PV) coupling versions
of the CBM, they found that the two are almost identical 
for the M1 transition,
except for the seagull term, which actually spoils 
the good fit to the experimental data. 
The recoil effects were neglected in both calculations, which were 
limited to photoproduction at the peak of the delta resonance.

In the standard treatment of the cloudy bag model, the cavity approximation 
for quark fields is adopted, and the pion field is then added in a 
perturbative way. 
The wave function of a bare baryon is  just a direct product of individual
quark wave functions, similar to the  nuclear shell-model 
wave function (independent particle motion).
This type of wave function is not a momentum eigenstate although
the Hamiltonian commutes with the total momentum operator.
The matrix elements evaluated between such static states contain 
spurious center of mass 
motion which ought to be removed.
Early studies indicated that the  correction for spurious 
 center of mass motion is significant\cite{Wilets}.
It is expected to be most important in calculations where  relatively
large momentum transfers are involved.
There are several intuitively motivated prescriptions\cite{Wilets}
 for the correction of
center of mass motion, however, none of them are fully satisfactory.
In this work, we have chosen to use  the  Peierls-Thouless (PT)\cite{PT62} 
method to eliminate the center of mass motion, since it is the most
convenient for our purposes. 
It is basically a nonrelativistic approach and 
requires two consecutive projections of the product wave functions to form
a Galilean invariant state for a composite baryon.
The energy expectation value evaluated using the PT projected  wave function
 has  the  correct nonrelativistic form, whereas this is not the case
for Peierls-Yoccoz projection.

In this paper, we calculate the nucleon-delta electromagnetic 
transition amplitudes
with respect to the virtual photon. The spurious center of mass
motion is corrected by using the PT projection method. 
As a first step, we assume exact $SU(6)$ symmetry for the quark structure 
of the baryons, so that all
quarks in the ground state of $N$ and $\Delta$ are in the $s$ state.
The notation of references\cite{TT83,THESIS} is followed.
We briefly review  the method to construct
 the PT wave function in Sec.\ II.\@ 
The calculation of helicity amplitudes for
virtual photoproduction of the delta is performed in Sec.\ III, 
and in Sec.\ IV we present the numerical results. Finally in Sec.\ V we
summarize our results.
 

\section{Galilean invariant baryon states}
We start with the chirally invariant Lagrangian density 
of the cloudy bag model\cite{CBM}
\begin{eqnarray}
        \protect{\cal L}
        &=&  (i\overline q \gamma^\mu \partial_\mu q - B)\theta_V 
        - {1\over 2}\overline q q \Delta_S \nonumber \\
	&& + {1\over 2} (\partial_\mu \bfgreek{\pi})^2 
        - {1\over 2} m^2_\pi \bfgreek{\pi}^2
	- {i\over 2f} \overline q \gamma_5 \bfgreek{\tau} \cdot 
	\bfgreek{\pi} q \Delta_S, \label{L}
\end{eqnarray}
where $\theta_V$ is a step function which is one inside the bag volume V 
and vanishes  outside, and
$\Delta_S$ is  a surface delta function.
In a perturbative treatment of the pion field, the quark wave function is not 
affected by the pion field and is given by the MIT bag solution\cite{MIT}
\begin{equation}
q({\bf r}) =  \left ( 
\begin{array}{c} g(r) \\ i\bfgreek {\sigma} \cdot\hat{r} f(r) \end{array} \right )  \theta(R-r),
\end{equation}
where $R$ is the spherical bag radius. 
For the ground state of a massless quark 
$g(r) = N_s j_0(\omega_s r/R), f(r) = N_s j_1(\omega_s r/R) $, 
where  $\omega_s = 2.0428$ and 
 $N_s^2 = \omega_s/8\pi R^3 j_0^2(\omega_s) (\omega_s -1)$.  

The  bare baryon is taken to be composed of three quarks with the spin-flavor  
wave function  given by $SU(6)$ symmetry. The space
component is naively the  direct product of three
quark wave functions in coordinate space
\begin{equation}
\Psi({\bf x}_1, {\bf x}_2, {\bf x}_3; {\bf x}) = 
q({\bf x}_1 - {\bf x}) q({\bf x}_2 - {\bf x}) q({\bf x}_3- {\bf x}).
\end{equation}
Here  ${\bf x}$ indicates the location of the bag center, while 
${\bf x}_1$, ${\bf x}_2$, and ${\bf x}_3$ specify the positions 
of the three quarks. Clearly this wavefunction 
does not have definite momentum and is 
not a momentum eigenstate. A momentum eigenstate 
of the  baryon can be constructed 
by making a linear superposition of the localized states, namely,
\begin{equation}
\Psi_{\rm{PY}}({\bf x}_1, {\bf x}_2, {\bf x}_3; {\bf p})
= N'(p) \int\! d^3{\bf x} e^{i {\bf p\cdot x}}
\Psi({\bf x}_1, {\bf x}_2, {\bf x}_3; {\bf x}),
\end{equation}
where the subscript PY stands for Peierls-Yoccoz projection\cite{PY57}, 
and $N'(p)$ is a momentum dependent normalization constant.
It can be shown that $\Psi_{\rm{PY}}({\bf p}) =
e^{i{\bf p\cdot x}_{\rm{cm}}} \Psi_{\rm{in}}({\bf p})$, where 
${\bf x}_{\rm{cm}} = ({\bf x}_1 + {\bf x}_2 + {\bf x}_3)/3$ 
is the the center of mass of the baryon
and $\Psi_{\rm{in}}({\bf p})$ is the appropriately defined intrinsic part of 
the wave function. 
Since $\Psi_{\rm{in}}({\bf p})$  still depends on the c.m. momentum, 
it violates  translational invariance. To overcome this problem, 
Peierls and Thouless (PT)\cite{PT62} proposed
to make another superposition of these momentum eigenstates, i.e.,
\begin{equation}
\Psi_{\rm{PT}}({\bf x}_1, {\bf x}_2, {\bf x}_3; {\bf p})
= N(p) \int\! d^3p' w({\bf p'}) e^{i({\bf p - p'}) \cdot {\bf x}_{\rm{cm}} }
\Psi_{\rm{PY}}({\bf x}_1, {\bf x}_2, {\bf x}_3; {\bf p'}).
\end{equation}
The weight function, $w(p')$,  should in fact be chosen to minimize
the total energy, but this would be quite complicated to implement here.
Instead, we choose $w({\bf p'}) = 1$ for simplicity
and convenience.  
Then integrations over  ${\bf x}$ and ${\bf p'}$ can be carried out easily.
This leads to a much simplified PT wave function,
\begin{equation}
\Psi_{\rm{PT}}({\bf x}_1, {\bf x}_2, {\bf x}_3; {\bf p}) = 
N_{\rm{PT}} e^{i{\bf p \cdot x}_{\rm{cm}}}
q({\bf x}_1 - {\bf x}_{\rm{cm}}) q({\bf x}_2 - {\bf x}_{\rm{cm}})
q({\bf x}_3 - {\bf x}_{\rm{cm}}), \label{PT}
\end{equation}
where the normalization factor, $N_{\rm{PT}}$, is given by the condition
\begin{equation}
\int\! d^3 x_1 d^3 x_2 d^3 x_3  
\Psi^\dagger_{\rm{PT}}({\bf x}_1, {\bf x}_2, {\bf x}_3; {\bf p'})
\Psi_{\rm{PT}}        ({\bf x}_1, {\bf x}_2, {\bf x}_3; {\bf p}) 
= (2\pi)^3 \delta^{(3)}(\bf p' - \bf p).
\end{equation}
This leads to 
\begin{equation}
N_{\rm{PT}} = \left[ 3\int\! d^3 r_1 d^3 r_2 \rho({\bf r}_1) \rho({\bf r}_2) 
\rho(-{\bf r}_1 - {\bf r}_2) \right ]^{-1/2},
\end{equation}
where $\rho({\bf r})$, the quark density at the position ${\bf r}$, 
is defined as 
\begin{equation}
\rho({\bf r}) \equiv q^\dagger({\bf r}) q({\bf r}) = [g^2(r) + f^2(r)]\, \theta(R-r).
\end{equation}
Notice that, for this simple version of the PT projection,
$N_{\rm{PT}}$ is a momentum independent constant and 
the wavefunction in Eq.\ (\ref{PT}) is manifestly Galilean invariant. 

\section{The helicity amplitudes in the cloudy bag model}

From the CBM Lagrangian given in  Eq.\ (\ref{L}), the conserved local  
electromagnetic current can be derived using 
 the principle of minimal coupling 
$\partial_\mu \rightarrow \partial_\mu + i q A_\mu$,
where $q$ is the charge carried by the field upon
which the derivative operator acts. 
The total electromagnetic current is then
\begin{eqnarray}
J^\mu(x) &=& j^\mu_q(x) + j^\mu_\pi(x), \\
j^\mu_q(x) &=& \sum_a Q_a e \overline{q}_a(x) \gamma^\mu q_a(x), \\
j^\mu_\pi(x) &=& -i e [ \pi^\dagger(x) \partial^\mu \pi(x)
               -\pi(x) \partial^\mu \pi^\dagger(x)],
\end{eqnarray}
where $q_a(x)$ is the quark field operator 
for flavor $a$, $Q_a$ is its charge in units of e, and $e \equiv |e|$ 
is the magnitude of the electron charge.
The charged pion field operator is defined as
\begin{equation} 
 \pi(x) = {1\over \sqrt{2}}[\pi_1(x) + i\pi_2(x)],
\end{equation}
where $\pi(x)$ either destroys a negatively charged pion 
or creates a positively charged one.

It is customary to define the helicity amplitudes for the electroproduction
of the delta as\cite{MUK87} 
\begin{eqnarray}
A_{3/2} &=& {1\over \sqrt{2 \omega_\gamma}}
\bra {\Delta; s_\Delta =3/2}  \vec{J}(0)\cdot \vec{\epsilon}
 \ket {N; s_N = 1/2}, \label{A3} \\
A_{1/2} &=& {1\over \sqrt{2 \omega_\gamma}}
\bra {\Delta; s_\Delta =1/2}  \vec{J}(0)\cdot \vec{\epsilon}
 \ket {N; s_N = -1/2}, \label{A1}
\end{eqnarray}
where the $\Delta$ is at rest and the photon is travelling along the z-axis
with 
right-handed 
polarization,  $\vec{\epsilon} = -{1\over\sqrt{2}} (1, i, 0)$.
The spin projections of $\Delta$ and $N$ 
along the z-axis are denoted as $s_\Delta$ and $s_N$ respectively.
For a virtual photon,
 the three-momentum in the $\Delta$ rest frame is given by 
\begin{equation}
|\vec{q}|^2 = Q^2 + {(M_\Delta^2 - M_N^2 - Q^2)^2 \over 4 M_\Delta^2},
\end{equation}
with $Q = \sqrt{-q^2}$ the magnitude of the four momentum transfer.
The photon energy is related to this by $q^2_0 = |\vec{q}|^2 - Q^2 $,
where for a real photon we have $Q^2 = 0$, so that 
$\omega_\gamma = |q_0| = (M_\Delta^2 - M_N^2)/2M_\Delta$.

The experimentally extracted, resonant, helicity amplitudes 
are to be associated 
with the fully dressed initial and final baryons. 
In the cloudy bag model, due to the $\pi BB'$ coupling, a physical baryon
state is  described as a  mixture of a bare bag and its surrounding pion cloud,
\begin{equation}
\ket A = \sqrt{Z^A_2} [ 1 + (E_A - H_0 - \Lambda H_I \Lambda )^{-1} H_I ] 
\ket {A_0}, \label{ST}
\end{equation}
where $Z^A_2$ is the bare baryon probability in the physical baryon states,
$\Lambda$ is a projection operator which projects out all the components
of $\ket A $ with at least one pion, and $H_I$ is the interaction Hamiltonian
which describes the process of emission and absorption of pions. 
The matrix elements of $H_I$ between the bare baryon states and 
their properties are then given by\cite{TT83,THESIS}
\begin{eqnarray}
v^{AB}_{0j}(\vec{k}) &\equiv& 
\bra {A_0} H_I \ket{{\bf \pi}_j(\vec{k}) B_0} = {i f^{AB}_0\over m_\pi}
{u(kR) \over [2\omega_k (2\pi)^3]^{1/2}} \sum_{m,n}
C^{s_B m s_A}_{S_B 1 S_A} (\hat{s}^*_m \cdot {\vec k}) 
C^{t_B n t_A}_{T_B 1 T_A} (\hat{t}^*_n \cdot {\vec e}_j),\\
w^{AB}_{0j}(\vec{k}) &\equiv& 
\bra{A_0 {\bf \pi}_j(\vec{k})} H_I \ket {B_0}
 = \left[v^{BA}_{0j}(\vec{k})\right]^* 
= -v^{AB}_{0j}(\vec{k}) = v^{AB}_{0j}(-\vec{k}),
\end{eqnarray}
where the pion has momentum $\vec{k}$ and  isospin projection $j$,
 $f_0^{AB}$ is the reduced matrix element for the 
$\pi B_0 \rightarrow A_0$ transition vertex, $u(kR) = 3j_1(kR)/kR $,
$\omega_k = \sqrt{k^2+ m^2_\pi}$, and $\hat{s}_m$ and $\hat{t}_n$ 
are spherical unit vectors for spin and isospin, respectively.

Under the approximation  that there is 
at most one pion in the air, there are three different
processes contributing to the $\gamma^* N \rightarrow\Delta$ vertex, 
as shown in Fig.\ 1.
Substituting the quark current operator, Eq.\ (11), 
and the physical baryon states, Eq.\ (17), into
Eqs.\ (\ref{A3}) and (\ref{A1}), we have
\begin{eqnarray}
&&\bra{\Delta,s_\Delta} \vec{j}_q\cdot\vec{\epsilon}\ket{N,s_N}
=\sqrt{Z^N_2 Z^\Delta_2} \left[
\bra{\Delta_0,s_\Delta} \vec{j}_q \cdot\vec{\epsilon}\ket{N_0,s_N}\right.
  \nonumber \\
&& \left. +\bra{\Delta_0,s_\Delta} H_I (E - H_0 - \Lambda H_I\Lambda)^{-1}
\vec{j}_q \cdot\vec{\epsilon}\,(E - H_0 - \Lambda H_I\Lambda)^{-1 }H_I
\ket{N_0,s_N}\right]
\end{eqnarray}
where $E$ is the total energy of this transition process. 
The first term, as illustrated in Fig.\ 1(a), is the quark core 
contribution and the second term, corresponding to Fig.\ 1(b), is the 
contribution from the $\gamma qq$ interaction with one $\pi$ in the air.
By inserting two complete sets of states before and after the operator
$\vec{j}_q$ in the second term and 
using the PT wavefunctions given in Eq.\ (\ref{PT}), we
obtain  
\begin{eqnarray}
A^{(a)}_{3/2}(Q^2) &=& \sqrt{3} A^{(a)}_{1/2}(Q^2) =  
A_{\rm{bare}}(Q^2) \sqrt{Z^N_2 Z^\Delta_2}, \\
A^{(b)}_{3/2}(Q^2) &=& \sqrt{3} A^{(b)}_{1/2}(Q^2) = 
A_{\rm{bare}}(Q^2) {(f^{NN})^2 \over 27\pi^2m_\pi^2} 
\int\! {dk k^4 u^2(kR)\over \omega_k} \left[
  {5/4 \over \omega_k (\omega_k + \delta - \omega_\gamma)}\right. \nonumber \\
&&\left.+ {1 \over (\omega_k + \delta)(\omega_k + \delta - \omega _\gamma)}
+ {2/25 \over (\omega_k + \delta)(\omega_k - \omega_\gamma)}
+ {1 \over \omega_k (\omega_k - \omega_\gamma)} \right], \label{AB}
\end{eqnarray}
where $\delta = m_\Delta - m_N$, and  $f^{NN}$ is the renormalized $\pi NN$
coupling constant. 
The four terms in the right-hand side of Eq.\ (\ref{AB})
correspond to four possible intermediate states, $(N\Delta)$, $(\Delta\Delta)$,
$(\Delta N)$, and $(N N)$, respectively. 
The recoil corrected bare $\gamma N_0\rightarrow\Delta_0$ 
transition amplitude is
\begin{equation}
A_{\rm{bare}}(Q^2) = -{ e\over \pi\sqrt{6\omega_\gamma}} 
{\int_0^R\! dr r^2 g(r) f(r) j_1(qr) K(r)
\over \int_0^R\! dr r^2 \rho(r) K(r)},
\end{equation}
where $K(r) = \int d^3x \rho(\vec{x}) \rho(-\vec{x} - \vec{r})$
 is the recoil function to account for the correlation of the 
two spectator quarks. 
The renormalization constants, $Z^A_2$, are  determined by the normalization
condition for the physical baryon state, i.e. 
\begin{equation}
Z^A_2 = \left[ 1 - {\partial\over \partial E}\Sigma^A(E)\right]^{-1}_{E=m_A},
\end{equation}
where $\Sigma^A$, the self energy of the baryon $A$, is given by
\begin{equation}
\Sigma^A(E) = \sum_B \left({f^{AB}_0\over m_\pi}\right)^2 {1\over 12\pi^2}
\, \mbox{P}\!\int\! {dk\, k^4 u^2(k R)\over \omega_k (E - m_B - \omega_k)},
\end{equation}
where P indicates that the principle part of the integral is to be taken.
In this work, we have adopted the usual philosophy for the renormalization
in the CBM. Throughout this work approximate relation,
$ f^{AB} \simeq \left({f^{AB}_0\over f^{NN}_0}\right) f^{NN}$, is always used.
There are uncertain corrections on the bare coupling contant $f^{NN}_0$,
such as the nonzero quark mass and correction of center of mass motion.
Therefore, we use the renormalized coupling constant in our
calculation, $f^{NN} \simeq 3.03$, 
which correspond to the usual $\pi NN$ coupling constant, 
$f^2_{\pi NN}\simeq 0.081$. 
As a result, the factor $\sqrt{Z^N_2 Z^\Delta_2}$ is absorbed into the 
renormalized coupling constants in Fig.\ 1(b). This treatment is equivalent
to the original CBM formalism
up to order $(f^{NN})^2$ and consistent with current conservation.

To evaluate the contribution caused by 
the photon-pion-pion coupling vertex [see Fig.\ 1(c)],
we use the usual plane wave expansion for the quantized pion field
\begin{equation}
\pi_j(\vec{x},t=0) = \int\! {d^3k\over [(2\pi)^3 2\omega_k]^{1/2}}
\left[ a_j(\vec{k}) e^{i\vec{k}\cdot\vec{x}} + 
a^\dagger_j(\vec{k}) e^{-i\vec{k}\cdot\vec{x}}\right],
\end{equation}
where $a_j(\vec{k})$  $(a^\dagger_j(\vec{k}))$ annihilates (creates) a pion
with momentum $\vec{k}$ and isospin $j$. The pion current operator becomes
\begin{equation}
\vec{j}_\pi(\vec{x}) = -i e \sum_{j j'} \epsilon_{jj' 3}
\int\! {d^3k d^3k'\over (2\pi)^3 2(\omega_k\omega_{k'})^{1/2}} 
      \, \vec{k}\,
e^{i (\vec{k} - \vec{k'})\cdot \vec{x}} 
\left[ a_{j'}(-\vec{k'}) + a^\dagger_{j'}(\vec{k'})\right]\cdot
\left[ a_j(\vec{k}) + a^\dagger_j(-\vec{k})\right].
\end{equation}
The transition amplitude at position $\vec{x}$ is just
the matrix element of $\vec{j}_\pi(\vec{x})$  evaluated 
between the physical baryon states. Using the identities
\begin{eqnarray}
a_j(\vec{k})\ket A &=& (E_A - \omega_k - H)^{-1} V^\dagger_{0j}(\vec{k})
\ket A, \\
a_{j'}(\vec{k'}) a_j(\vec{k})\ket A &=& 
(E_A - \omega_k - \omega_{k'} - H)^{-1} V^\dagger_{0j}(\vec{k})
(E_A - \omega_{k'} - H)^{-1} V^\dagger_{0j'}(\vec{k'}) \nonumber \\
&+& 
(E_A - \omega_k - \omega_{k'} - H)^{-1} V^\dagger_{0j'}(\vec{k'})
(E_A - \omega_{k} - H)^{-1} V^\dagger_{0j}(\vec{k}) \ket A,
\end{eqnarray}
we have
\begin{eqnarray}
\bra{\Delta,s_\Delta} \vec{j}_\pi(\vec{x}) \ket{N,s_N} 
&=& -i e \sum_{j j'}\epsilon_{jj' 3} \int\! d^3k d^3k' 
e^{i (\vec{k} - \vec{k'})\cdot \vec{x}} 
{\vec{k} \over (2\pi)^3 2(\omega_k\omega_{k'})^{1/2}} \nonumber \\
&\times&
\sum_B \left[ \eta^B_{j'j}(\vec{k'},\vec{k}) G^B(\vec{k'},\vec{k}) +
               \eta^B_{jj'}(\vec{k},\vec{k'}) G^B(\vec{k},\vec{k'})\right].
\end{eqnarray}
Here $B$ denotes the intermediate baryon states 
(restricted to $N$ and $\Delta$ here), and 
\begin{eqnarray}
\eta^B_{j'j}(\vec{k'},\vec{k}) &\equiv& {f^{\Delta B} f^{NB} \over m^2_\pi}
{u(kR) u(k'R) \over 16\pi^3 (\omega_k \omega_{k'})^{1/2}}\nonumber\\
&\times& \sum_{s_B} C^{s_B m' s_\Delta}_{S_B 1 S_\Delta} C^{s_B m s_N}_{S_B 1 S_N}
             (\hat{s}^*_{m'} \cdot {\bf k'}) (\hat{s}^*_{m} \cdot {\bf k})
   \sum_{t_B} C^{t_B n' t_\Delta}_{T_B 1 T_\Delta} C^{t_B n t_N}_{T_B 1 T_N}
             (\hat{t}^*_{n'} \cdot {\bf e}_j)(\hat{t}^*_{n} \cdot {\bf e}_j),\\
G^N(\vec{k'},\vec{k}) &\equiv& 
  {1\over (\omega_k + \omega_{k'} + \delta)\omega_{k}}
+ {1\over (\omega_{k'} -\omega_\gamma)\omega_{k}}
+ {1\over (\omega_{k'} -\omega_\gamma)(\omega_k + \omega_{k'} -\omega_\gamma)},
\label{gn} \\
G^{\Delta}(\vec{k'},\vec{k}) &\equiv& 
  {1\over (\omega_k + \omega_{k'} + \delta)(\omega_k + \delta)}
+ {1\over (\omega_{k'} + \delta - \omega_\gamma)(\omega_k + \delta)}\nonumber\\
&&
+ {1\over (\omega_{k'} + \delta - \omega_\gamma)
 (\omega_k + \omega_{k'} - \omega_\gamma)}. \label{gd}
\end{eqnarray}
$G^{\Delta}(\vec{k},\vec{k'})$ and $G^{\Delta}(\vec{k},\vec{k'})$ are obtained
by the interchange of $\vec{k}$ and $\vec{k'}$ in the corresponding equation. 
 The three terms appearing in Eqs.~(\ref{gn}) and (\ref{gd}) 
correspond to the three
different time orders in the time-ordered perturbation theory, as illustrated
in Fig.\ 1(c).
Using the translational invariance of the electromagnetic current operator,
$j^\mu(x) = e^{i\hat{p}\cdot x} j^\mu(0) e^{-i\hat{p}\cdot x}$, then
the $\gamma^* N\rightarrow\Delta$  helicity amplitudes due to the 
$\gamma \pi\pi$ interaction are simply given by 
\begin{equation}
A(Q^2) = \int\! d^3x e^{i \vec{q}\cdot\vec{x}}
\bra{\Delta,s_\Delta} \vec{j}_\pi(\vec{x})\cdot \vec{\epsilon}\ket{N,s_N}.
\end{equation}
After performing some  spin and isospin  algebra, we obtain
\begin{eqnarray}
A^{(c)}_{3/2}(Q^2) &=&  
- {(f^{NN})^2 |\vec{q}| \over 240\sqrt{6\omega_\gamma}\pi^3 m_\pi^2} 
\int\! {d^3k k^4 \sin^2\!\theta \, u(kR) u(k'R)\over  \omega_k \omega_{k'}}
\nonumber\\
&\times& \left[ G^N(\vec{k}, \vec{k'}) + 
3 G^\Delta(\vec{k'}, \vec{k}) + 2 G^\Delta(\vec{k},\vec{k'} )\right], \\
A^{(c)}_{1/2}(Q^2) &=&  
- {(f^{NN})^2 |\vec{q}| \over 720\sqrt{2\omega_\gamma}\pi^3 m_\pi^2} 
\int\! {d^3k k^4 \sin^2\!\theta \, u(kR) u(k'R)\over  \omega_k \omega_{k'}}
\nonumber\\
&\times& \left[ 2 G^N(\vec{k'}, \vec{k}) - G^N(\vec{k}, \vec{k'}) + 
 G^\Delta(\vec{k'}, \vec{k}) + 4 G^\Delta(\vec{k},\vec{k'} )\right],
\end{eqnarray}
where 
$\vec{k'} = \vec{k} + \vec{q}$, $\omega_{k'} = \sqrt{k'^2 + m^2_\pi}$,
and $\theta$ denotes the angle between $\vec{k}$ and $\vec{q}$.
It is worthwhile to  mention that the form of our results for Fig.\ 1(c) 
are expressed in a quite different
form than those of KE\cite{KE83} and Bermuth et al.\cite{Bermuth86} where
the integral variables are $k$ and $k'$ in their formulation. We believe that 
our expressions are more straightforward and manifestly  
respect the three momentum conservation 
at the $\gamma\pi\pi$ vertex. 
Our numerical results for this contribution also
appear to differ from those of Refs.\cite{KE83,Bermuth86}.

\section{Numerical Results}

Figure 2 shows the recoil function $K(r)$ due to the two quark spectators. 
It cuts down  the contributions from
the quark wave functions near the bag boundary by  up to 50 \%. 
The overall effect of this recoil
correction on the bare bag contribution to the 
typical $\gamma^* N\rightarrow\Delta$ helicity amplitude,
 $A_{3/2}$, 
is shown in Fig.~\ref{fig.bare}. 
In the real photon limit ($Q^2 \rightarrow 0$), 
the magnitude of the $\gamma^* N\rightarrow\Delta$ transition
amplitude increases with the bag radii in a fashion similar to 
that of the magnetic moment
of bare baryons. The correction of the center of mass motion usually reduces
 the bare transition
amplitudes by 5 to 10 \% for $Q^2 \alt 0.5 \mbox{ GeV}^2$ within a reasonable
range of bag radii. However, this recoil correction would flip sign and
 increase the transition amplitude for larger momentum transfers.


The expressions of  $A^{(b)}$ and $A^{(c)}$ 
(from Fig.\ 1(b) and 1(c) respectively) involve pion-loop integrals,
which contain poles originating from the baryon mass difference 
in the propagators.  
Using the relation P $ \int^\infty_0 {dk\over k^2 - k^2_0} = 0$, 
it can be shown that
\begin{equation}
 \lim_{\epsilon \rightarrow 0^+} 
\int^\infty_0\! {dk f(k)\over k^2 - k^2_0 - i\epsilon} = 
 \int^\infty_0\! dk {f(k)-f(k_0)\over k^2 - k^2_0} + {i\pi\over 2 k_0} f(k_0),
\end{equation}
where $k_0 > 0$.
To be consistent with the phenomenology of the cloudy bag model,
 we limit our calculation to the small momentum
transfer region (approximately $Q^2 \alt 1.6 \mbox{ GeV}^2$).
 For  higher momentum transfers,  relativistic effects are essential and   
  the support of the $\theta$ integral is considerably more complicated.
Consequently, for our purposes,
 there is only one relevant pole, $\omega_k = \omega_\gamma$, for  Fig.\ 1(b) 
and an extra pole, $\omega_{k'} = \omega_\gamma$, for Fig.\ 1(c).


Fig.~\ref{fig.part}  shows the individual contributions 
to the helicity amplitude, $A_{3/2}$, from the various diagrams for a typical
bag radius, R = 0.8 fm.
It is clear that the contributions from the pion cloud are significant
in the cloudy bag model. 
As $Q^2 \rightarrow 0$, the pionic effects account for two thirds of the total
amplitude, or roughly twice
that of the quark core for the
chosen bag radius. As $Q^2$ increases beyond $0.2 \mbox{ GeV}^2$, 
the contribution from the $\gamma\pi\pi$ interaction decreases rather 
rapidly, so that it has nearly vanished at $Q^2 \sim \mbox{ 1.2 GeV}^2$. 
For $Q^2 \agt 1 \mbox{ GeV}^2$, the pionic effects contribute
 about 40 \% of the total amplitude.
Notice that, since the $\Delta$ is a resonance, 
the helicity amplitudes are actually complex.
In the CBM the imaginary part arises from  thresholds in the 
pion loop integral, and amounts to $15 \sim 20 \% $ of the real part.

The real parts of total helicity amplitudes, 
$A = A^{(a)} + A^{(b)} + A^{(c)}$, 
are presented
in Fig.~\ref{fig.helicity}. With the contributions of the pion cloud,
the bag radius dependence is quite different from that for the bare transition
amplitudes shown in Fig.\ 3. This can be explained
by the fact that 
the pionic contribution is competing with that of  quark core,
since a small bag radius means a strong pion cloud. 
In the small $Q^2$ region, the pion cloud compensates more 
than the loss in the bare amplitude when using a 
small bag radius. 
Generally, the smaller the bag radius, 
the larger the total transition amplitude.
We list the helicity amplitudes corresponding to the  real photon limit
at the $\Delta$ resonance  in Table 1.
With the small bag radius  $R$ = 0.7 fm, 
we are able to reproduce the experimental helicity
amplitude in this model.

Finally, in Fig.~\ref{fig.m1}, 
we show the real part of the invariant $\gamma^* N\rightarrow\Delta$ 
magnetic transition form factor, $G^M_\Delta(Q^2)$,
in comparison  with the experimental measurements\cite{DATA}.
The relation between the helicity amplitudes 
and the invariant transition form factors are given 
in Ref.\cite{WARNS90}.
With the bag radius $R = 1 \mbox{ fm}$, 
our calculation agrees reasonably well with the data
for modest $Q^2$, but is less satisfactory 
at  $Q^2 \simeq 0 \mbox{ GeV}^2$ and 
in the region $Q^2 \agt 1 \mbox{ GeV}^2$. At the real photon limit, 
we are able to
get the PDG\cite{PDG} value with $R = 0.7 \mbox{ fm}$, 
however, at nonzero $Q^2$ the prediction with this bag radius would not
be in agreement with the data. 
The  center of mass correction  reduces
$G^M_\Delta(0)$ by approximately 5 \% and 
makes the M1 form factor somewhat harder,
similar to what was found in the case of the  
nucleon magnetic form factor\cite{preprint}.

In the present work the quark cores of nucleon and delta  were assumed
to have  SU(6) symmetry,
i.e., 
there is no deformation assumed, so that the spatial parts of 
the wave functions
for the $N$ and $\Delta$ are purely $s$ state. 
Hence, here the only possible source 
for an  $E2$ amplitude is
from Fig.\ 1(c). Using Eqs.\ (21), (22), (35) and (36), we obtain
$E2/M1 = (A_{1/2}-A_{3/2}/\sqrt{3})/(A_{1/2}+\sqrt{3}A_{3/2}) 
\sim -0.3 \times 10^{-3}$.  
 Because of the severe cancellation
in the numerator, this result is about two orders of magnitude smaller
than the phenomenological approach\cite{DAVID91,PDG} 
(The latest estimate by
 Particle Data Group is $-0.015 \pm 0.004$).

\section{summary}
In conclusion, we have calculated the 
$\gamma^* N \rightarrow\Delta$ transition 
form factors in the cloudy bag model, 
including the center of mass correction via 
a simplified Peierls-Thouless projection method.
The effect of this recoil correction is to slightly 
reduce the magnetic form factor
at small momentum transfer and to  
make the form factor slightly  harder. Generally, with the PT projection
the transition moment is reduced by about $5 \sim 8 \%$.

The pion cloud contribution proved to be crucial to account for 
the measured helicity amplitudes using a reasonable bag radius in this model.
In similar calculations using constituent quark models 
(the nonrelativistic\cite{IK82} and relativized quark models\cite{WARNS90}), 
 the helicity amplitudes
are usually significantly underpredicted. 

Because of the large range of momentum transfer measured, we have been 
unable to obtain
a totally satisfactory description of all of the data using  our 
essentially nonrelativistic
formalism. Typically the form factors are too stiff. 
 Unfortunately, there is as yet no covariant formalism
for quark bag models or soliton bag models.  
Using a relatively simple prescription
for Lorentz boosts\cite{LP}, the invariant form factor $G^M_\Delta (Q^2)$
has shown improved behaviour. Obviously a  crucial next step will be to 
construct  a quark
model with improved Lorentz transformation properties, in addition to
the inclusion of a perturbative pion cloud.
It will also be important to explore the effect of $SU(6)$ violating admixtures
in the baryon wave functions\cite{IK82}.

This work was supported by the Australian Research Council.

\begin{table}
\caption{Helicity amplitude of delta photoproduction, $A_{3/2}$, 
in units of $10^{-3} \mbox{GeV}^{-1/2}$.
Here static denotes the static calculation and PT denotes the Peierls-Thouless
projection. The indices a, b, and c 
correspond to the Figs.~1(a), 1(b), and 1(c) respectively. 
The latest estimate by Particle Data Group is $-258 \pm 6$. }
\begin{center}
\begin{tabular}{l|cccc|cccr}
 & \multicolumn{4}{c|}{static}   &  \multicolumn{4}{c}{PT}  \\ \hline
  R(fm)& a & b & c & total & a & b & c & total \\ \hline
 1.0 &-115 & -53 -21i & -64 -18i & -205 &-106 & -49 -19i & -64 -18i & -195\\
 0.9 &-106 & -62 -20i & -80 -20i & -216 & -98 & -58 -18i & -80 -20i & -205\\
 0.8 & -96 & -73 -19i &-101 -21i & -233 & -86 & -68 -17i &-101 -21i & -222\\
 0.7 & -85 & -87 -17i &-129 -22i & -260 & -79 & -80 -16i &-129 -22i & -249
\end{tabular}
\end{center}
\end{table}

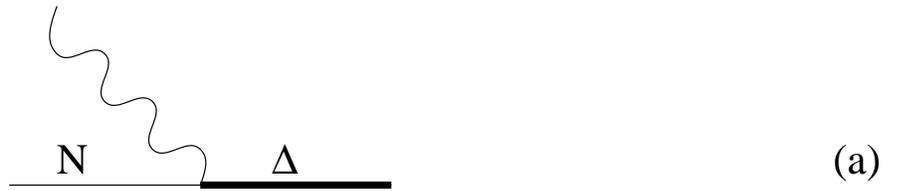
\begin{figure}
\caption{Diagrams illustrating the various contributions included 
in the calculation. The intermediate baryons $B$ and $B'$ are
restricted to the 
$N$ and $\Delta$ here.}
\label{fig1}
\end{figure}

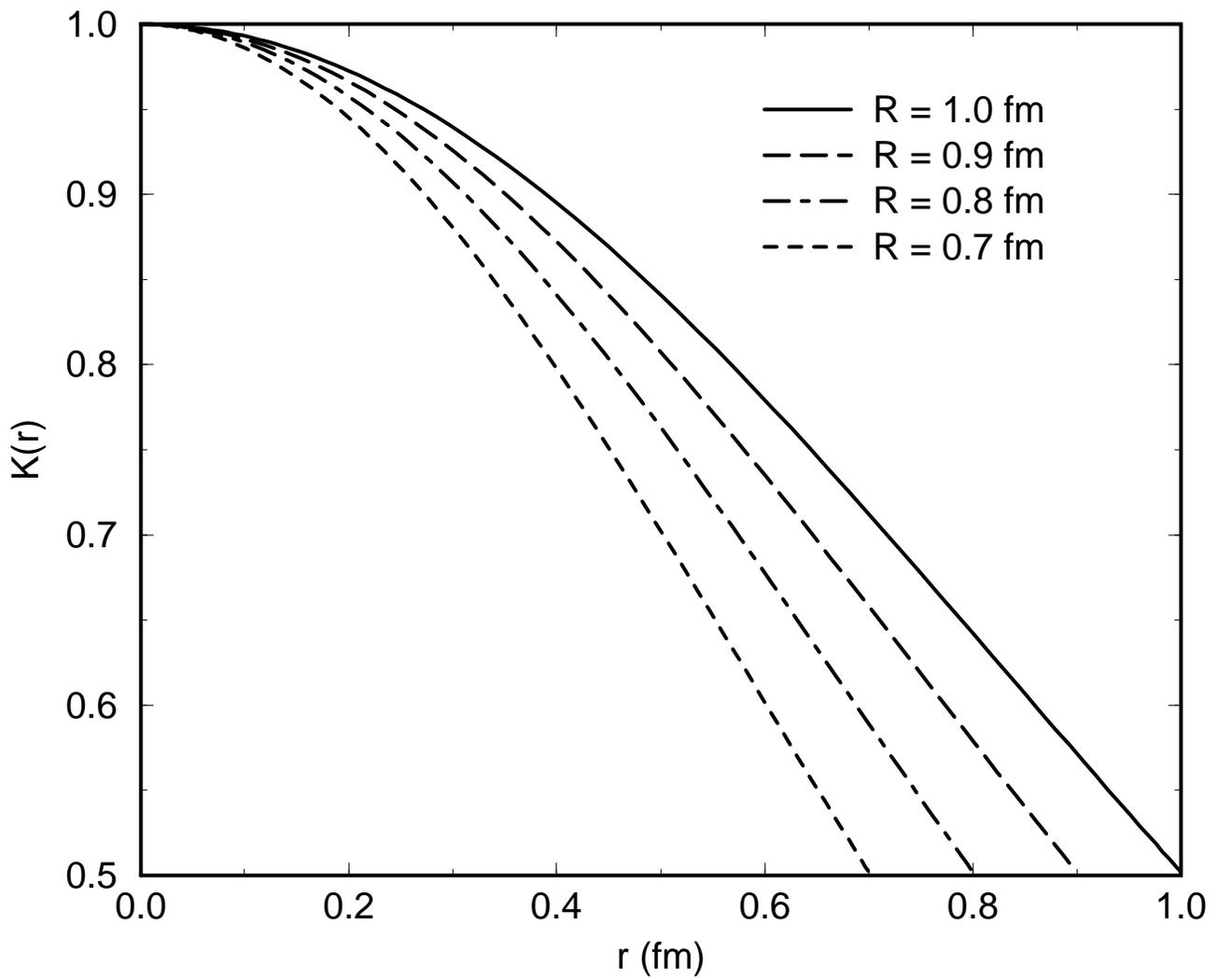
\begin{figure}
\caption{The recoil function K(r), defined following Eq.\ (23),
 due to the two quark spectators
in  arbitrary units.} 
\label{fig2}
\end{figure}

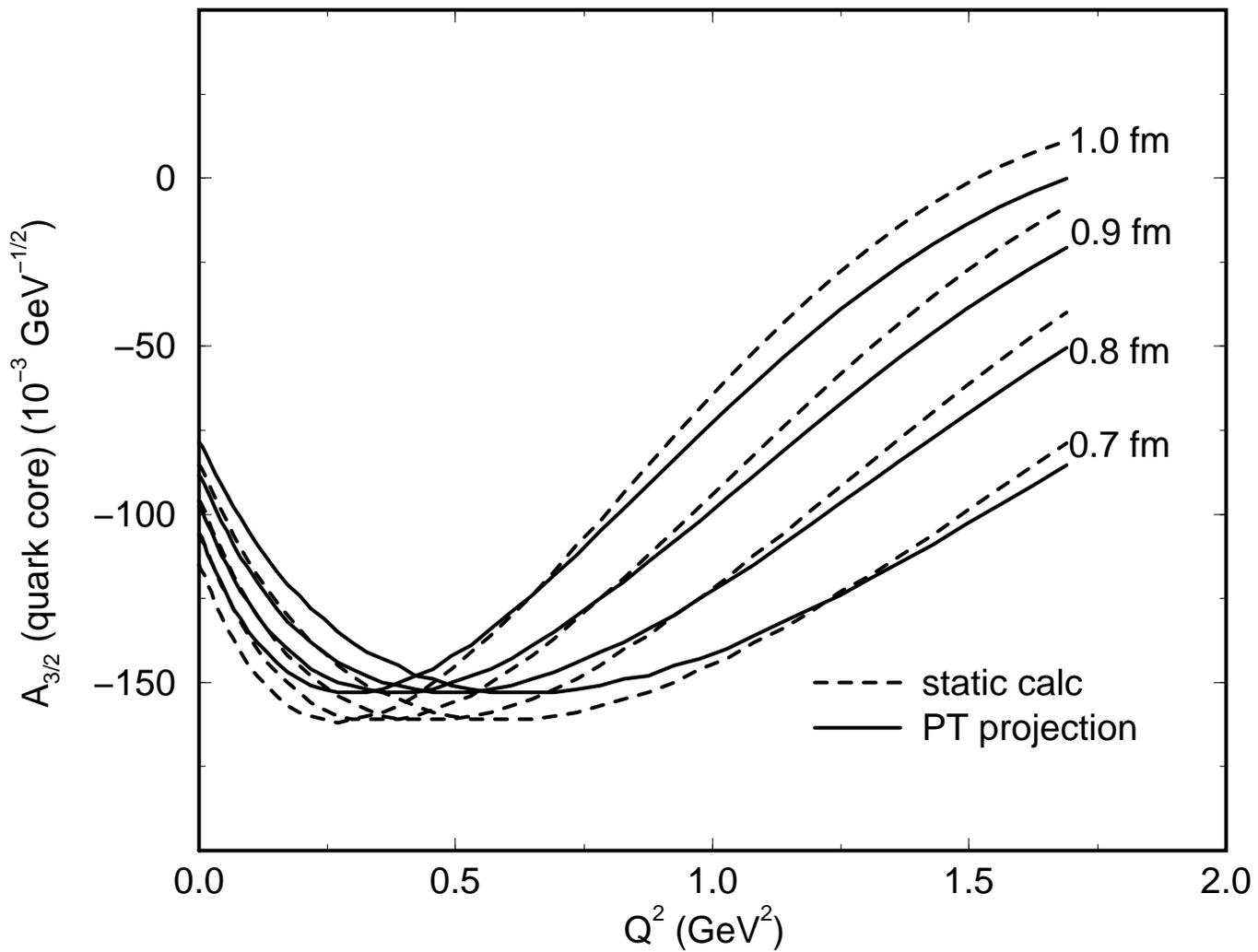
\begin{figure}
\caption{The effect of the center of mass correction on the 
helicity amplitude, $A_{3/2}$, for the bare bag. The number 
on each curve 
indicates the bag radius in fm  for the calculation.}
\label{fig.bare}
\end{figure}

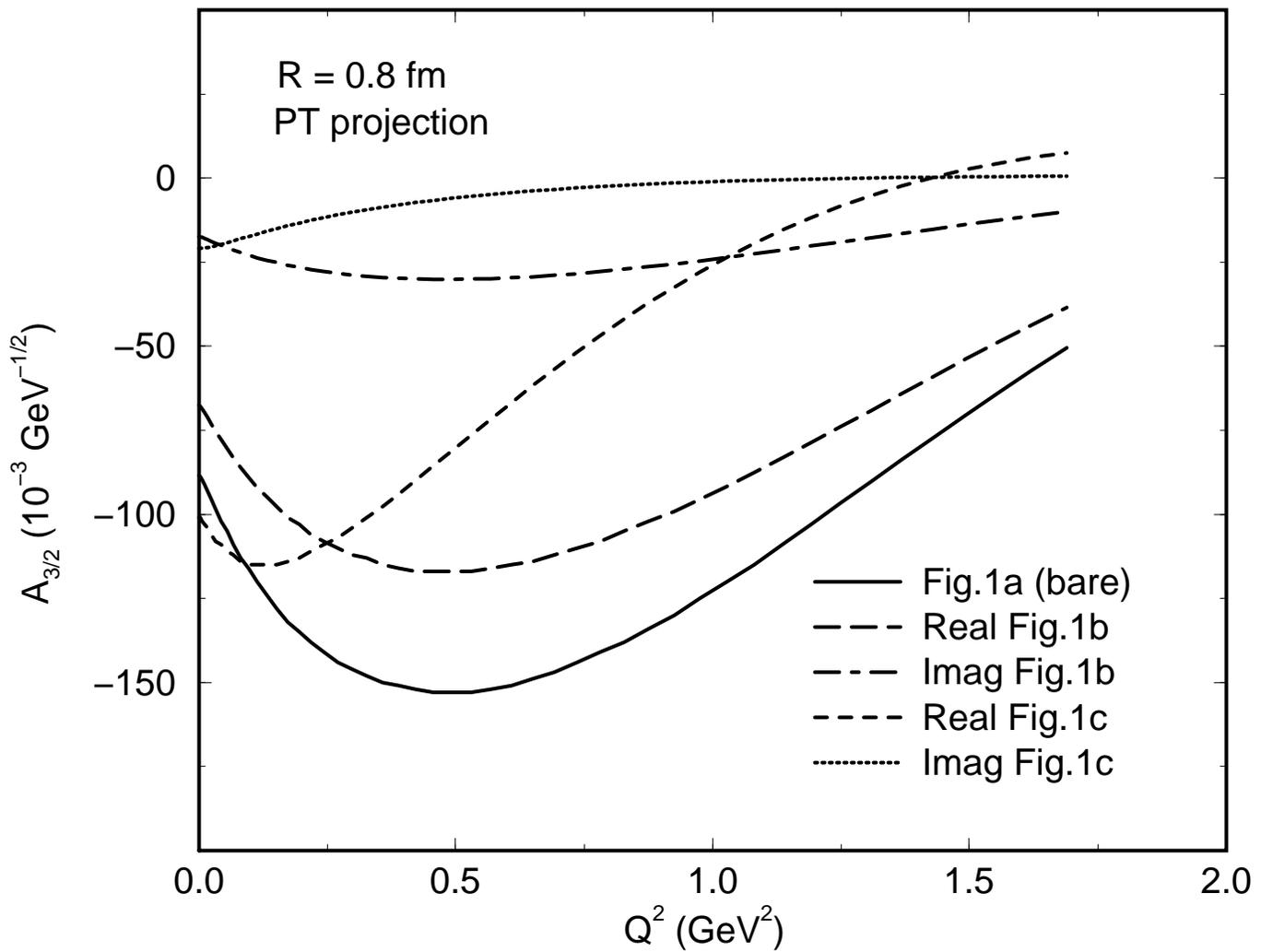
\begin{figure}
\caption{ Individual components for the helicity amplitude, $A_{3/2}$,
 with a typical bag radius R = 0.8 fm.}
\label{fig.part}
\end{figure}

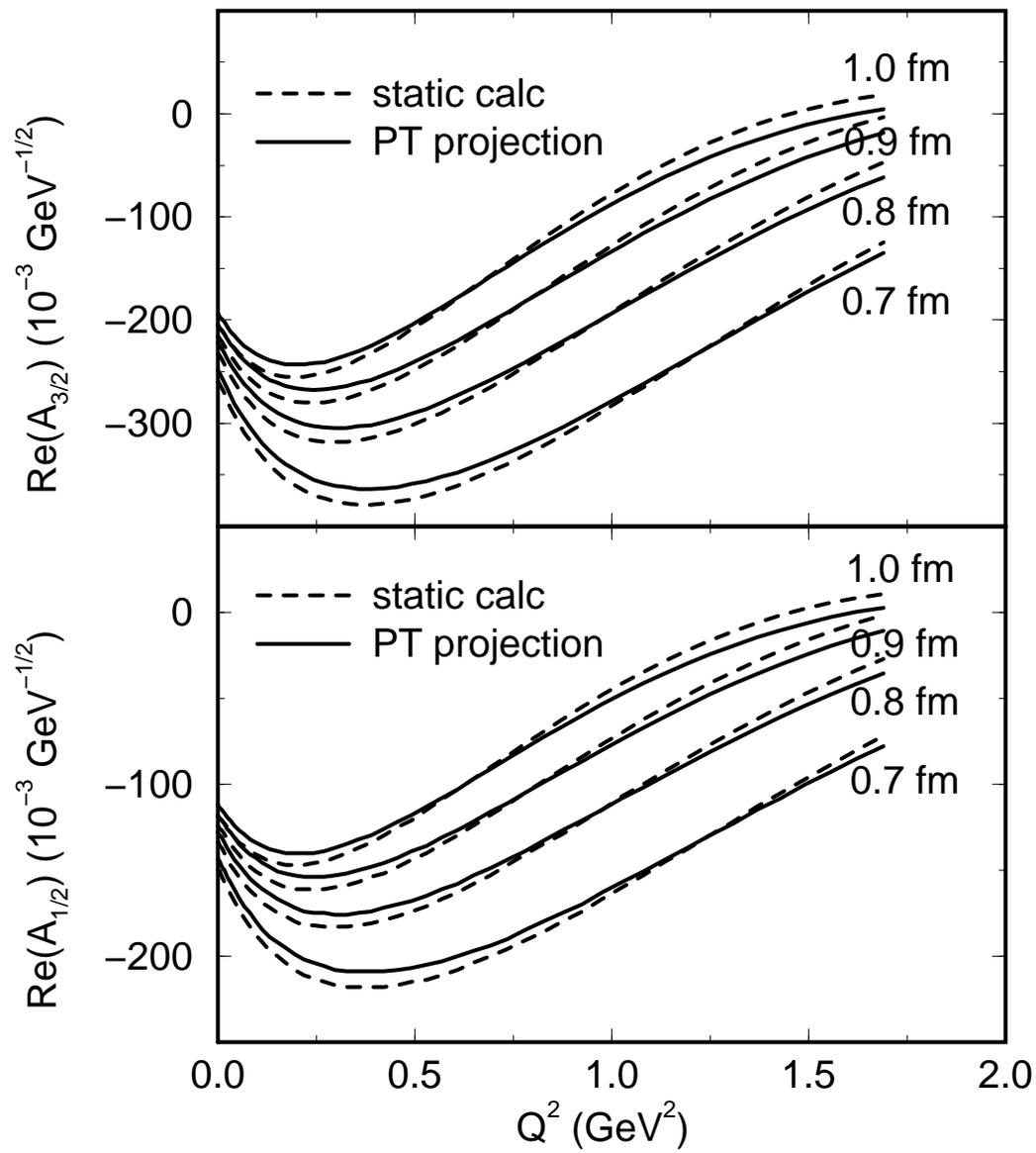
\begin{figure}
\caption{ The real parts 
of total $\gamma^*N\rightarrow\Delta$ helicity amplitudes, 
Re[$A_{3/2}(Q^2)$] and Re[$A_{1/2}(Q^2)$].}
\label{fig.helicity}
\end{figure}

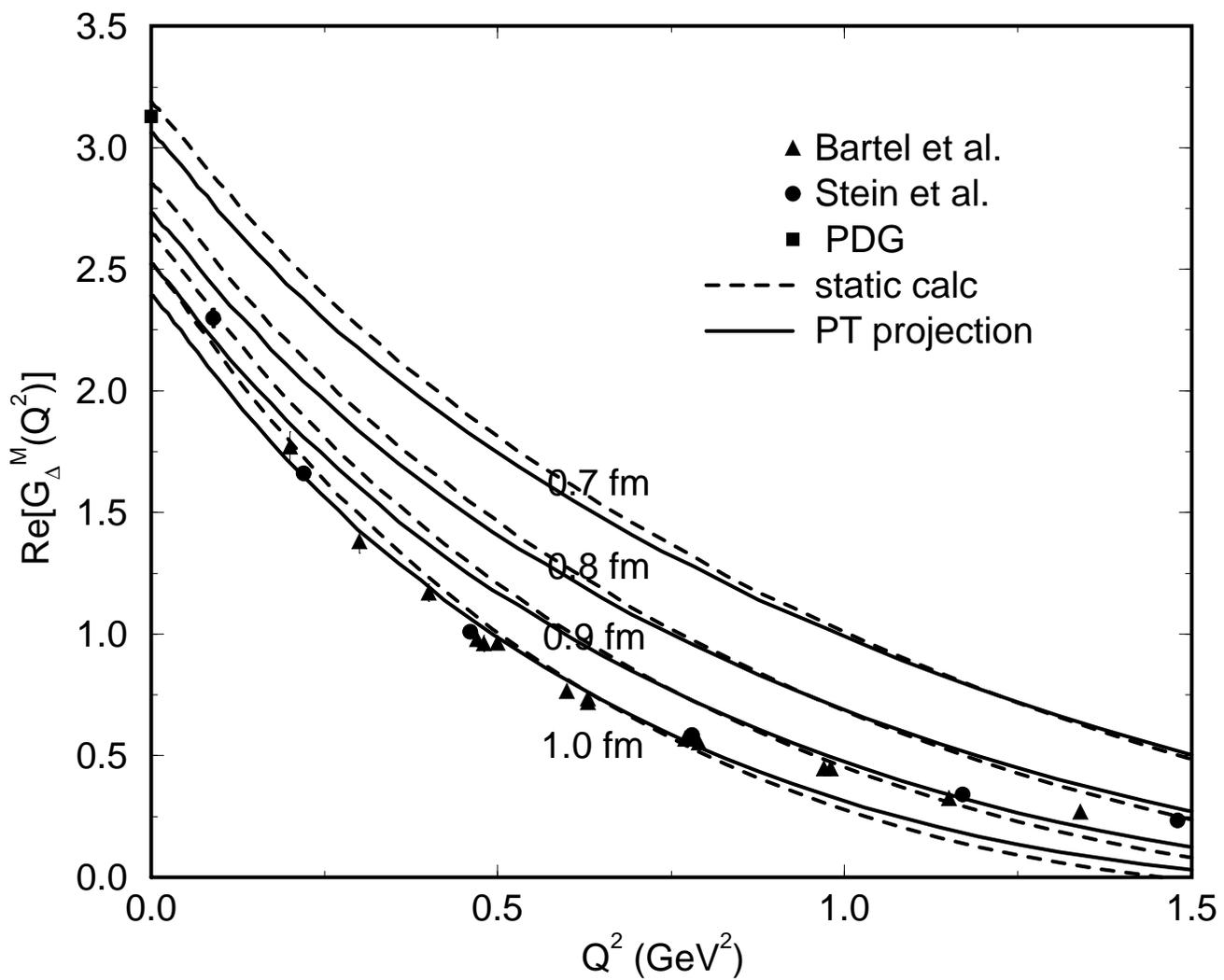
\begin{figure}
\caption{ The real part of 
$\gamma^*N\rightarrow\Delta$ invariant M1 transition 
form factor, Re[$G^M_\Delta(Q^2)]$. Four different bag radii are used 
in the calculation. 
The experimental data are taken from Refs.\ [16,17].}
\label{fig.m1}
\end{figure}

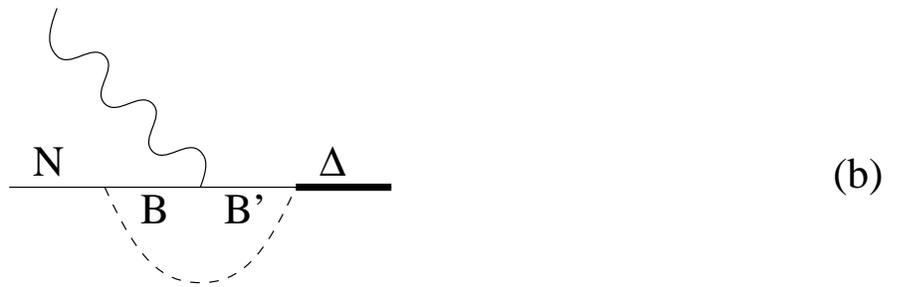
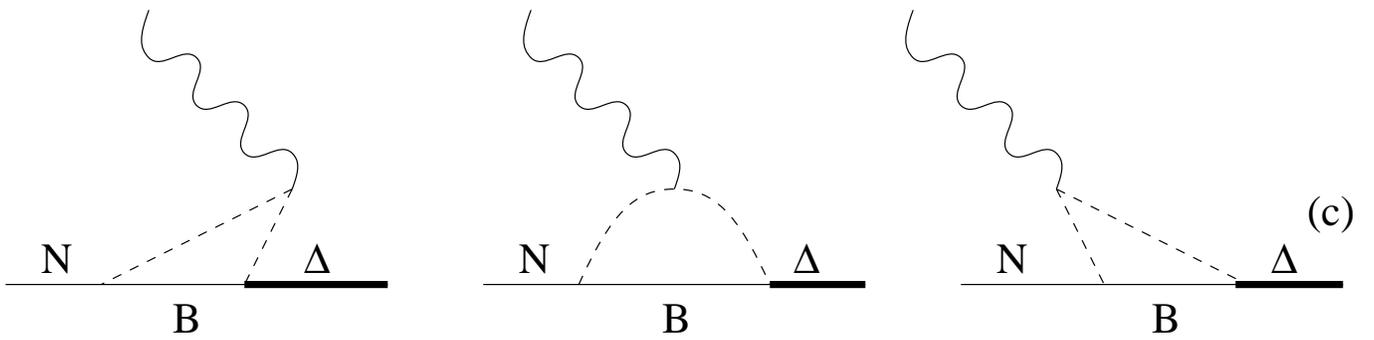
\end{document}